# SINGLE-SHOT ELECTRO-OPTIC SAMPLING OF COHERENT TRANSITION RADIATION AT THE A0 PHOTOINJECTOR

T. J. Maxwell[1,2], J. Ruan[2], P. Piot[1,2] and R. Thurman-Keup[2]
[1] Department of Physics, Northern Illinois University, DeKalb, IL 60115, USA
[2] Fermi National Accelerator Laboratory, Batavia, IL 60510, USA

*Abstract*

Future collider applications and present high-gradient laser plasma wakefield accelerators operating with picosecond bunch durations place a higher demand on the time resolution of bunch distribution diagnostics. This demand has led to significant advancements in the field of electro-optic sampling over the past ten years. These methods allow the probing of diagnostic light such as coherent transition radiation [1] or the bunch wakefields [2] with sub-picosecond time resolution. Potential applications in shot-to-shot, non-interceptive diagnostics continue to be pursued for live beam monitoring of collider and pump-probe experiments. Related to our developing work with electro-optic imaging, we present results on single-shot electro-optic sampling of the coherent transition radiation from bunches generated at the A0 photoinjector.

## INTRODUCTION

We report on single-shot, electro-optic spectral encoding [1, 3, 4, 5] of THz-regime coherent transition radiation (CTR) from the A0 photoinjector for longitudinal bunch length measurements.

The ultimate goal is to probe the time-resolved transverse bunch distribution in a plane [6]. For more advanced 3D designs [7], techniques relying on spatial encoding become problematic as spatial encoding requires non-collinear mixing of the signal and a laser stripe in the transverse plane to introduce a known delay.

To begin we detail the beam and laser probe used and present our approach to matching the time of arrival of the CTR to the probe. We then explain a balanced spectral EOS detection scheme. Bunch length data is presented with a comparison to measurements by a Martin-Puplett interferometer [8].

Discussion of spectral encoding is done in the most basic approximation, though it has been shown that spectral encoding introduces artifacts in this case [9] except under restrictive conditions [10].

## BEAM, PROBE, AND TIMING

The probe laser is a 1-kHz repetition rate Spitfire Pro XP regenerative amplifier seeded by a Tsunami laser oscillator (Spectra-Physics) with an 800-nm center wavelength. The system is also equipped with a longitudinal pulse shaper (DAZZLER by FASTLITE). The amplifier was initially chosen as probe for its pulse duration controls, large available pulse energy, and for the easy isolation of the 1-Hz pulse for experiment.

Generation of short electron bunches was done using the emittance-exchange (EEX) line at the A0 photoinjector [11]. When using the EEX line, the input transverse phase space is mapped to the longitudinal. Therefore we can minimize the bunch length after EEX by adjusting the strength of the quads upstream of the EEX line. The output beam is focused by the downstream quads onto an aluminium mirror oriented 45° to beam's propagation direction at diagnostic cross "X24."

Transition radiation from the mirror is collimated by a parabolic reflector and can be directed to the EOS experiment, a synchronized streak camera [12], pyroelectric detector, or the interferometer [8].

The laser pulse is sent from the A0 laser lab to the accelerator tunnel through the existing optical transport line. It's then split from the UV and imaged to an optical breadboard installed at X24 (Figure 1) where EOS of the beam CTR is performed.

Synchronization of the short laser and CTR pulses is accomplished in three steps: For the ns-scale, time of arrival of the IR probe laser and UV drive laser at the photoinjector's RF gun are observed on a photodiode. The IR laser phase is set to arrive 3.5 ns earlier than the UV to account for the path difference between the e-beam and probe laser travelling from the RF gun to X24.

At the breadboard, an alternate optical path (not shown) then combines laser leakage and the optical transition radiation, sending it to the streak camera for timing to tens of picoseconds. Accounting for an additional ~260 ps path difference, the EOS signal is then monitored as the probe delay is scanned over a ~20-ps range until a signal is observed.

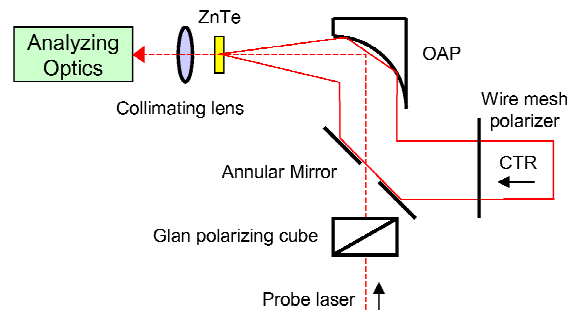

Figure 1: EOS breadboard layout. Probe laser (dashed red) and CTR (solid red) combined and focused on ZnTe.

## BALANCED SPECTRAL ENCODING

In the spectral encoding scheme, a chirped laser pulse propagates collinearly with the CTR through a 1 mm ZnTe crystal (Figure 1). To ensure a well-defined interaction in the polarization-sensitive crystal, the laser polarizer is set to admit light with polarization perpendicular to that of the admitted CTR.

In the most basic approximation, the ZnTe crystal appears to the laser as an ultra-fast Pockels cell with retardance proportional to the strength of the CTR [13]. With a strong linear chirp on the laser pulse, the time dependence of the retardance is mapped to a spectral amplitude modulation.

The modulated laser pulse is sent to the analyzing optics (Figure 2). For sign-resolved measurements, the modulated signal passes through a quarter wave plate with fast axis 45° from the laser polarizer shown in Figure 1. This produces circular polarization of the laser in the absence of the ZnTe crystal.

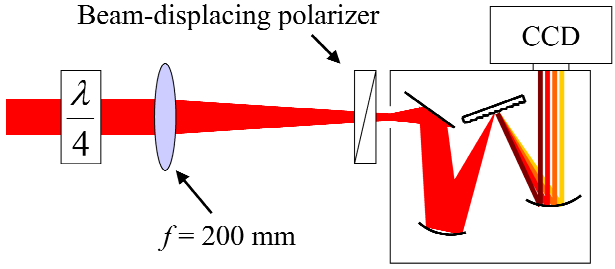

Figure 2: EOS spectral analyzing optics detailing the polarization-resolving spectrometer configuration. Beam displacement occurs 3 mm into the plane shown, parallel to the spectrometer slit.

The signal is then sent to a polarization-resolving spectrometer [14]. As the laser is focused onto the vertical slit of the spectrometer, it passes a beam-displacing polarizer that produces two parallel beams. The undeviated beam is the horizontal component of the incident polarization while the vertical component is displaced 3 mm vertically. The two beams traverse the spectrometer in parallel, producing two horizontal stripes in the image plane with the top (bottom) stripe resolving the spectrum of the horizontal (vertical) polarization.

The spectrometer is a Princeton Instruments SP-2150 equipped with a 600 g/mm grating blazed at 750 nm. A 16-bit, 768x1024 CCD camera mounted in the image plane is used as detector. For typical focusing and slit width, the spectral resolution is found to be 0.15 nm. The polarization-dependent response of the system is measured independently with subsequent measurements corrected accordingly.

Following a similar description as [3, 13], we approximate the temporal modulation of the chirped laser pulse as an equivalent spectral modulation mapped by the instantaneous wavelength ($\lambda = \lambda_{inst}(t)$), which is set by the chirp of the laser. In the time domain, the difference of intensities for the two polarizations after the quarter wave plate is found to be

$$I_v[\lambda(t)] - I_h[\lambda(t)] = I_{laser}(t)\sin[\Gamma(t)] \qquad (1)$$

Here $I_{laser}$ is the unmodulated laser intensity and $\Gamma(t)$ the induced retardance, in this case proportional to the field of the CTR, $E_{CTR}(t)$. After normalizing by the measured, unmodulated laser intensity, and knowing the laser chirp such that the map $\lambda(t)$ can be computed, a sign-resolved estimate of $E_{CTR}(t)$ is straightforward.

For a complete map of wavelength to time, longitudinal laser phase reconstruction is accomplished by second harmonic generation, frequency-resolved optical gating (SHG FROG) [15].

## RESULTS

Single-shot measurements of the CTR emitted from short, 250-pC electron bunches were performed at the A0 photoinjector. As determined by SHG FROG, the chirped laser pulse length is 4.4 ps FWHM. In the current approximation, for chirped pulse length $\tau_c$ and Fourier-limited length of $\tau_o$, this sets a temporal resolution of $\sqrt{\tau_c \tau_0}$ = 660 fs, FWHM [3, 9, 10, 13]. Recovered retardances, proportional to the CTR transient $E_{CTR}(t)$, are shown in Figure 3 for two different bunches.

From Figure 3 it is apparent that the oscillatory signals are not a direct indicator of the bunch distribution. This is due to a number of effects not accounted for including transport response of the CTR [16], phase mismatch and signal distortions in the thick crystal [17], and a more accurate treatment of the encoding process [4].

For comparison, we estimate the signal FWHM as half the width between extrema of one full cycle of the largest oscillation. For Figure 3, these occur around –1.8 ps and 0.8 ps in both sets. This results in durations of 1.44 ps for bunch 1 and 1.14 ps for bunch 2.

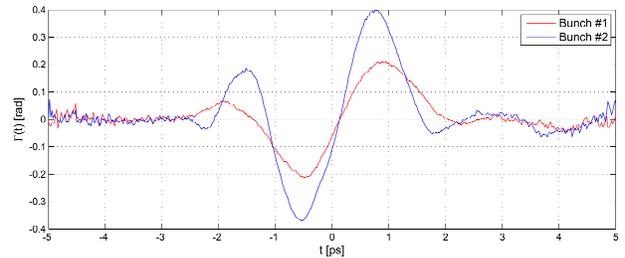

Figure 3: Recovered retardances induced by the CTR from 250 pC bunches of two different bunch lengths, estimated as 1.14 ps (blue) and 1.44 ps (red), FWHM.

Interferometry of $E_{CTR}(t)$ was independently performed in both cases using the Martin-Puplett interferometer [8]. Assuming $E_{CTR}(t)$ is real and proportional to the deduced retardances (Figure 3), the equivalent interferogram (IF) of the recovered EOS signal is

$$IF_{EOS}(\tau) \propto \int |\Gamma(t) + \Gamma(t-\tau)|^2 \, dt \qquad (2)$$

These are shown in Figure 4 for both bunches using only vertical scaling to equalize peak values of the IFs shown.

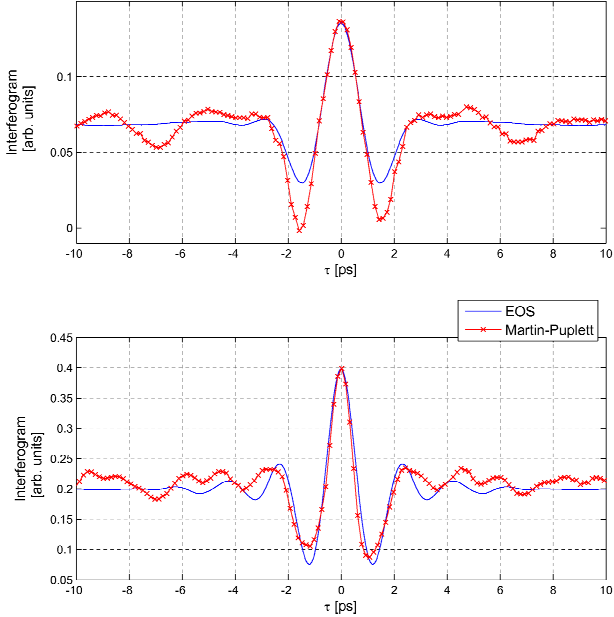

Figure 4: Interferogram data for bunch #1 (top) and bunch #2 (bottom) including the Martin-Puplett interferometer scans and IF of the EOS signals (eq. 2).

As above, we define the FWHM of the AC as half the distance between the strong minima (near ±1.5 ps). In this case, for the EOS signals, the IF width of bunches 1 and 2 are 1.48 ps and 1.21 ps, respectively. Corresponding interferometer scans are in agreement with widths of 1.50 ps and 1.17 ps, respectively.

Further, despite the interferometer more readily probing the long tail of the IF beyond the sensitivity and time window of the EOS measurement, as well as differences in system responses, in both cases similar features in the tail are observed.

## SUMMARY

The installation of EOS diagnostics at A0 photoinjector has been shown. Details on the general synchronization and balanced detection schemes used were also presented. Further, CTR signals being analyzed are in agreement with independent measurements.

Minimization of the probe pulse is important in reducing the interference of the signal being measured with the THz light generated by optical rectification of the laser in the crystal. We therefore note that this single-shot approach provided the demonstrated sensitivity using laser pulses with pulse energies of less than 20 nJ.

It is known that true reconstruction of the bunch distribution from the will require the consideration of other known effects. With spectral encoding suggesting an avenue for 2D and 3D bunch distribution diagnostics, analysis of data from this experiment continues in support of properly decoding measured signals.

## ACKNOWLEDGEMENTS

This work was supported by the Fermi Research Alliance, LLC under U.S. Dept. of Energy Contract No. DE-AC02-07CH11359, and Northern Illinois University under US Department of Defense DURIP program Contract N00014-08-1-1064.

We would like to thank group members Jayakar Charles Tobin, Alex Lumpkin, Amber Johnson, James Santucci, Yin-e Sun, Helen Edwards and Michael Church of Fermi National Accelerator Laboratory for helpful discussions and technical support.

## REFERENCES

[1] J. van Tilborg *et al.*, Phys. Rev. Lett. **96**, 014801 (2006).
[2] M. J. Fitch *et al.*, Phys. Rev. Lett. **87** 034801 (2001).
[3] Z. Jiang and X.-C. Zhang, Appl. Phys. Lett. **72**, 1945 (1998).
[4] K. Y. Kim *et al.*, Appl. Phys. Lett. **88**, 041123 (2006).
[5] J. van Tilborg *et al.*, Optics Letters **33**, 1186-1188 (2008).
[6] T. Maxwell and P. Piot, Proceedings of the IEEE Particle Accelerator Conference 2009, Vancouver, BC, Canada, 3968 (2009).
[7] H. Tomizawa, *et al.*, Proceedings of the Free Electron Laser Conference 2010, Malmö, Sweden, 445 (2010).
[8] R. Thurman-Keup, R.P. Fliller, and G. Kazakevich, Proceedings of the Beam Instrumentation Workshop 2008, Lake Tahoe, CA, 153-157 (2008).
[9] S. P. Jamison *et al.*, Opt. Lett. **28**, 1710 (2003).
[10] X.-Y. Peng, J.-H. Teng, X.-H. Zhang, and Y.-L. Foo, J. Appl. Phys. **108**, 093112 (2010).
[11] J. Ruan, A. Johnson, A. Lumpkin, R. Thurman-Keup, H. Edwards, R. Fliller, T. Koeth, and Y.-E Sun, Physical Review Letters **106**, 244801 (2011).
[12] A. H. Lumpkin and J. Ruan, Proceedings of the Beam Instrumentation Workshop 2008, Lake Tahoe, CA, 365-369 (2008).
[13] J. van Tilborg, "Coherent Terahertz Radiation from Laser-Wakefield-Accelerated Electron Beams", Ph.D Thesis, Eindhoven University of Technology (2006).
[14] J. Kim and D. Kim, Rev. Sci. Inst. **79**, 033109 (2008).
[15] R. Trebino, *Frequency-Resolved Optical Gating: The Measurement of Ultrashort Laser Pulses*, Kluwer Academic Publishers, Boston (2002).
[16] S. Casalbuoni, B. Schmidt and P. Schmüser, "Far-Infrared Transition and Diffraction Radiation," TESLA Report 2005-15 (2005).
[17] S. Casalbuoni *et al.*, "Numerical Studies on the Electro-Optic Sampling of Relativistic Electron Bunches," TESLA Report 2005-01 (2005).